\documentclass[acus]{JAC2000}

\usepackage{graphicx}

\def\subdef#1{\gdef\globalColor##1{##1}}
\def\newColor #1 {\expandafter\def\csname #1\endcsname##1{##1}
   \expandafter\def\csname text#1\endcsname{\subdef{#1}
    }}%
\newColor GreenYellow     
\newColor Yellow          
\newColor Goldenrod       
\newColor Dandelion       
\newColor Apricot         
\newColor Peach           
\newColor Melon           
\newColor YellowOrange    
\newColor Orange          
\newColor BurntOrange     
\newColor Bittersweet     
\newColor RedOrange       
\newColor Mahogany        
\newColor Maroon          
\newColor BrickRed        
\newColor Red             
\newColor OrangeRed       
\newColor RubineRed       
\newColor WildStrawberry  
\newColor Salmon          
\newColor CarnationPink   
\newColor Magenta         
\newColor VioletRed       
\newColor Rhodamine       
\newColor Mulberry        
\newColor RedViolet       
\newColor Fuchsia         
\newColor Lavender        
\newColor Thistle         
\newColor Orchid          
\newColor DarkOrchid      
\newColor Purple          
\newColor Plum            
\newColor Violet          
\newColor RoyalPurple     
\newColor BlueViolet      
\newColor Periwinkle      
\newColor CadetBlue       
\newColor CornflowerBlue  
\newColor MidnightBlue    
\newColor NavyBlue        
\newColor RoyalBlue       
\newColor Blue            
\newColor Cerulean        
\newColor Cyan            
\newColor ProcessBlue     
\newColor SkyBlue         
\newColor Turquoise       
\newColor TealBlue        
\newColor Aquamarine      
\newColor BlueGreen       
\newColor Emerald         
\newColor JungleGreen     
\newColor SeaGreen        
\newColor Green           
\newColor ForestGreen     
\newColor PineGreen       
\newColor LimeGreen       
\newColor YellowGreen     
\newColor SpringGreen     
\newColor OliveGreen      
\newColor RawSienna       
\newColor Sepia           
\newColor Brown           
\newColor Tan             
\newColor Gray            
\newColor Black           
\newColor White           

\subdef{Black}

\begin{document}
\title{\flushright{WEDT003}\\[15pt] \centering INTRODUCING I/O CHANNELS
  INTO THE DEVICE DATABASE OPENS NEW POTENTIALITIES FOR CONFIGURATION
  MANAGEMENT \thanks{Funded by the Bundesministerium f\"ur Bildung,
    Wissenschaft, Forschung und Technologie (BMBF) and the Land Berlin.}}

\author{T. Birke, B. Franksen, R. Lange, P. Laux, R. M\"uller, BESSY,
Berlin, Germany}

\maketitle

\begin{abstract}
The reference RDBMS for BESSY II has been set up with a device oriented
data model. This has proven adequate for e.g. template based RTDB
generation, modelling etc. But since assigned I/O channels have been stored
outside the database (a) numerous specific conditions had to be maintained
within the scripts generating configuration files and (b) several generic
applications could not be set up automatically by scripts. In a larger
re-design effort the I/O channels are introduced into the RDBMS. That
modification allows to generate a larger set of RTDBs, map specific
conditions into database relations and maintain application configurations
by relatively simple extraction scripts.
\end{abstract}

\section{Introduction}

Within the last few years {\em Relational DataBase Management Systems}
(RDBMS or DB in short) have become essential for control system
configuration. With availability of generic applications and hardware
standards the interplay of components as well as the adaptation to site
specific needs has become crucial. Proper networking has been the central
problem of the past.  Today's challenge is a central repository of
reference and configuration data as well as an appropriate standard suite
of DB applications. Target is a management system providing consistency in
a programmable, comprehensible and automatic way for development, test and
production phases of the control system. Instead of careful bookkeeping of
innumerable hand-edited files the ever needed modifications of the facility
require `only' change of atomic and unique configuration data in the DB and
eventually adaptation of structures in the DB (applications). The update of
tool configurations is then consistently accomplished by direct DB
connection, renewal of snapshot files generated by extraction scripts etc.

\section{Status and Achievements}

Very early a device oriented approach has been chosen for the description
of the 3rd generation light source BESSY~II. A naming convention was
developed for easy identification and parsing of classifying properties
like installation location, device family, type and instance. Around the
`bootstrap' information contained in the device names a first reference
database has been set up \cite{ica97}, describing wiring, calibrations,
geometries etc.

\subsection{Device I/O} 
Utilisation of the EPICS control system toolkit implicates the network
protocol called {\em Channel Access} for the I/O of process variables. At
BESSY the device model results in a scheme {\tt <DEVICENAME>:<channel>}.

Generation of the EPICS {\em Real Time DataBases} (RTDB) for device classes
with high multiplicity and simple I/O (power supplies, vacuum system,
timings) is based on two components: device class specific data are stored in
the DB.  Functionality and logic of {\tt <channel>} is modelled with a
graphical editor and stored in a template file. Generating scripts merge
both into the actual RTDB. 


\begin{figure}[hb]
\centering
\includegraphics*[width=70mm]{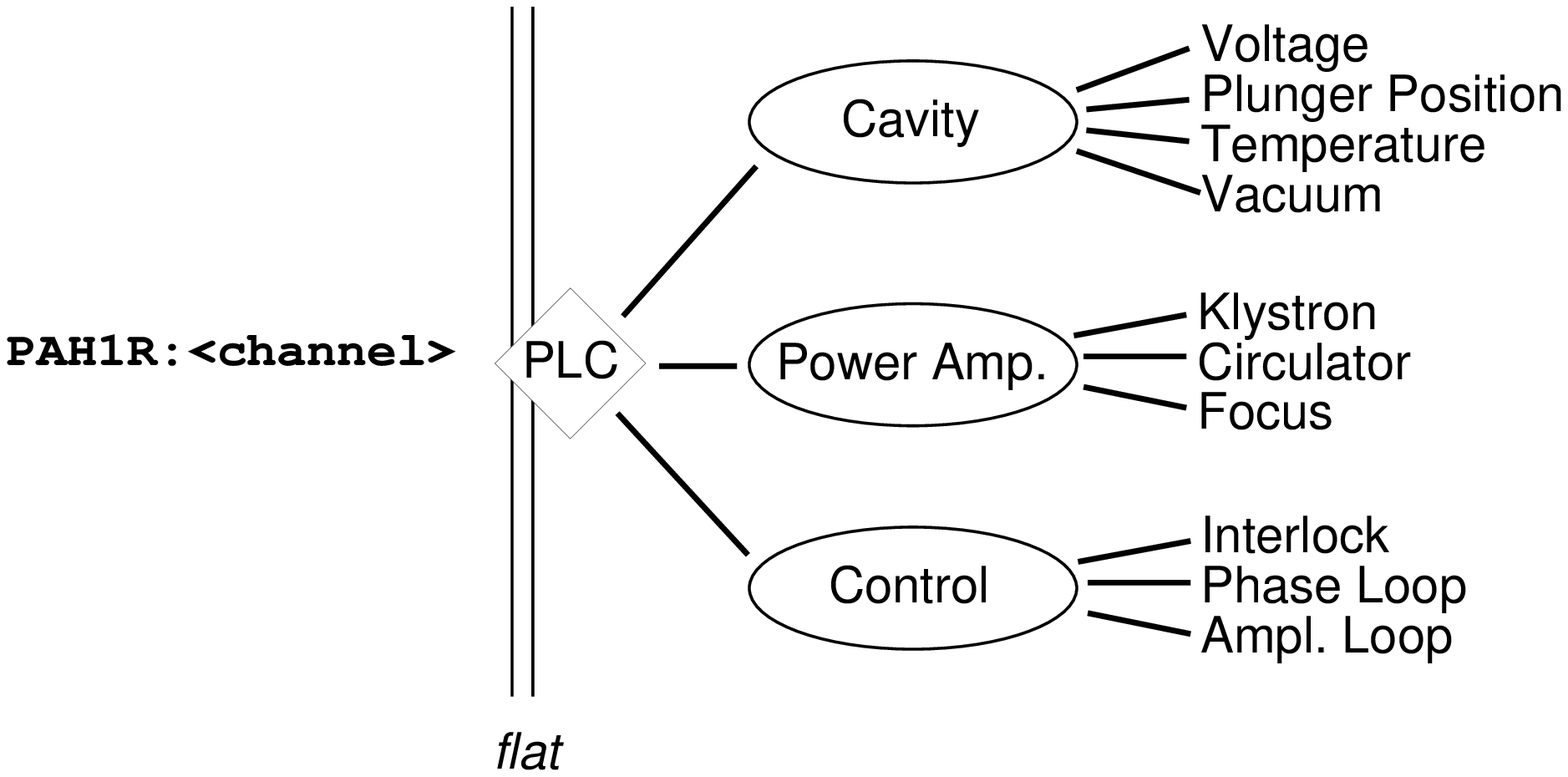}
\caption{RF Power Units are controlled by flat CAN I/O to complex,
  turn-key PLC systems. Addressing of sub-units is done via channel (name)
  grouping.}
\label{RFNew}
\end{figure}

For unique and complex systems (RF, insertion devices) \label{RF}
structuring takes place in {\tt <channel>} where no common naming
convention has been defined so far. Contrary to the previous approach of
complex template/atomic substitution here all channels involved are
assigned to sub-units and hierarchies.  The structure is fully implemented
in the DB and for RTDB generation only connected with simple atomic
templates.

Intermediate systems (scraper, GP-IB devices) are either adapted to one of
these approaches or simply set up by ad-hoc created files.

\subsection{Conditioning} 

Documentation of and transitions between different facility operation
conditions are handled by a save/restore/compare tool that is working on a
set of snapshot files. Coarse configuration is provided by SQL retrievals
of name lists. Hierarchies and partitioning are derived from device-name
patterns and mapped into directories and files. This is feasible only
because the relevant channels are restricted to setpoint, readback and
status.  Deviations of this scheme are few and easily maintainable by hand.

Information required by modelling tools for conversion between engineering
units of device I/O and the physics views (magnet function, length,
position, conversion factor etc.)  are fully available in the DB
\cite{ABS}. Therefore the linear optics correction tools (orbit, tune) are
instantaneously consistently configured provided the installed hardware 
matches the entries in the DB.

\subsection{Alarm, Archiving} 
As long as the alarm handler has to monitor only hardware trips DB
retrieval of device name collections and script based interpretation of
device name patterns are helpful at least for the simple devices with high
multiplicity: typical channels are ON/OFF status. Again grouping and
hierarchies can be derived from the class description embedded in the
device name.

For the complex devices (RF) control functionality and logic has already
been mapped into DB structures (see \ref{RF}). Here genuine DB calls
produce alarm handler configuration files with sophisticated error
reporting capabilities.

In a similar fashion creation of configuration files for the data collector
engine(s) of the archiving system is simplified by DB calls: for each
device class a limited number of signals and associated frequencies are of
interest for long term monitoring. The DB basically serves as source for
device name collections.

\subsection{Data Transfer} 
High level software using the CDEV API (configured with a {\em device
description language} file) and `foreign' networks (connected by the {\em
Channel Access gateway}) benefit also from the DB: For CDEV access
definition and permission to selected I/O channels for each device class is
defined by appropriate prototype descriptions (in analogy to the RTDB
templates). The associated lists of devices are compiled with DB calls.
{\em CA} gateway takes advantage of the naming convention by regular
expression evaluation.

\subsection{Installation Procedures} 
Facility modifications typically change the device inventory. The DB
supports consistent propagation of innovations \cite{ica99}: During an
installation campaign devices are added or deleted in the DB. In addition,
new device classes are modelled within script logics and template
(prototype) descriptions.  Running the configuration scripts on all control
system levels updates the configurations within development
environment, test system and production area.

\section{Solvable Deficiencies}
A number of restrictions imposed by the present device oriented DB model
are solvable by minor structural modifications and consequent introduction
of channels necessary for the adequate description of essential device
properties.

\subsection{Device Ambiguities, Hierarchies}
Dependent on the specific point of view, definition of device classes and
assignment of equipment comprises ambiguities. E.g. the device class {\em
magnet} ({\tt M}) is tightly bound to the model aspect of the storage ring,
whereas the the class {\em power supply} ({\tt P}) covers the engineering
aspects of the current converters. Pulsed elements (Kicker, Septa) are very
similar devices, but do not fit into this scheme - neither into the model
nor the I/O aspects. The decision to assign a single dedicated device-class
({\tt K}) introduces a new pattern (fig.~\ref{DevNew}) and breaks the
naming structure.

\newcommand{\HC}{\MidnightBlue}
\newcommand{\M}{\Magenta}
\newcommand{\R}{\Red}
\newcommand{\OR}{\Goldenrod}
\newcommand{\RO}{\RedOrange}
\newcommand{\G}{\PineGreen}
\newcommand{\B}{\NavyBlue}
\newcommand{\Y}{\Yellow}
\newcommand{\T}{\Turquoise}
\newcommand{\5}{\Gray}
\newcommand{\0}{\Black}
\newcommand{\RedBall}{\R{$\bullet$}\,}
\newcommand{\BB}{\0{$\bullet$}}

\begin{figure}[ht]
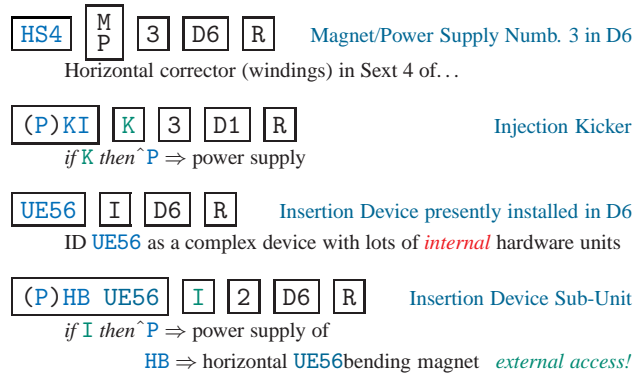

\begin{description}
\item[{\tt \fbox{\B{HS4}} \fbox{$\mbox{M}\atop\mbox{P}$} \fbox{3} \fbox{D6}
    \fbox{R}}] \hfill {\footnotesize \HC{Magnet/Power Supply Numb. 3 in D6}}\\
  {\footnotesize Horizontal corrector (windings) in Sext 4 of\ldots}
\item[{\tt \fbox{(\B{P})\B{KI}} \fbox{\G{K}} \fbox{3} \fbox{D1} \fbox{R}}]
  \hfill {\footnotesize \HC{Injection Kicker}}\\
  {\footnotesize {\em if} {\small\tt\G{K}} {\em then} {\small $\hat{}$ \tt\B{P}} $\Rightarrow$ power supply}
\item[{\tt \fbox{\B{UE56}} \fbox{I} \fbox{D6} \fbox{R}}] \hfill {\footnotesize
    \HC{Insertion Device presently installed in D6}}\\
  {\footnotesize ID {\small\tt\B{UE56}} as a complex device with lots of \R{{\em
        internal}} hardware units}
\item[{\tt \fbox{(\B{P})\B{HB} \HC{UE56}} \fbox{\G{I}} \fbox{2} \fbox{D6} \fbox{R}}] \hfill 
{\footnotesize \HC{Insertion Device Sub-Unit}}\\
  {\footnotesize {\em if} {\small\tt\G{I}} {\em then} {\small $\hat{}$ \tt\B{P}} $\Rightarrow$ power supply of\\
    \hspace*{1cm} {\small\tt\B{HB}} $\Rightarrow$ horizontal {\small\tt\HC{UE56}}bending
    magnet \hfill\G{{\em external access!}}}
\end{description}
\caption{Ambiguities in device class definition result in a complicated
  naming structure}
\label{DevNew}
\end{figure}

Difficulties get worse for complex pieces of equipment: insertion devices
have to be treated as `units'. They can be completely replaced by different
entities with similar complexity. Typically they consist of a number of
devices partly belonging to already described device classes (like bipolar
power supplies).  

The straightforward solution here is the extension of the
device naming convention: substructures and naming rules have to be
introduced also for the (up to now monolithic) device property describing
name element (`genome chart').  Corresponding DB structures have to be
implemented.

Similarly naming rules have to be developed where structuring takes place
in {\tt <channel>} (fig. \ref{RFNew}).  In summary, the distinction and
boundary between device and {\tt <channel>} are dictated by specifics of the
I/O connectivity and arbitrary with respect to the required DB structure.
It does not necessarily coincide with the most adequate classification
aspects. 

\subsection{Context Dependent Role of Devices}

Several global states are well defined for the whole facility (e.g.
`shutdown', `machine development', `user service' etc.) or for sub-sections
or device collections: `injection running', `wave length shifter ON' etc.
The role of devices depend on these states: in the context of
`shutdown' insufficient liquid helium level at super-conducting devices has
to generate a high severity alarm.  For power supplies even OFF states are
then no failure. But during `user service' alarms should notify the
operator already when power supply readbacks are out of meaningful bounds.

Similar case distinctions have to be made for all conditioning
applications (reload constraints for save/restore, active/inactive elements
for modelling), for data archiving (active periods, frequency/monitor
mode), for sequencing programs etc. Today only the most general conditions
are statically configured and available from the database/configuration
scripts. Exceptions are partly coded within the applications or have to be
handled by the operators --- an error-prone situation.

A clearly laid out man-machine interface provides an effective protection
against accidental maloperation. Presentation details and accessibility of
devices should be tailored to the user groups addressed: Operators, device
experts, accelerator physicists.  The required attributes attached to the
device channels should be easily retrievable from the DB.

\section{Structural Shortcomings}

\subsection{Immature Data Model}

Starting from the device model a very specific view of the control system
structure has been mapped into DB structure: device classes correspond to
tailored set of tables.  Deviating aspects that do not fit into the scheme
are modelled by relations, constraints, triggers and presented to the
applications by pre-built views. The result is not very flexible, hard to
extend and complicated to maintain in a consistent state free from
redundancies.  Numerous device attributes are scattered over
template/prototype files and configuration creating scripts in an implicit
format without clear and explicit relation to the data within the DB.

\subsection{Unavailable Data}

Configurations of archiver retrieval tools are much more demanding than
those of data collector processes. In search of correlations arbitrary
channel names have to be detected by their physical dimension: for a drift
analysis the data sources for temperatures [C], RF frequency [MHz], BPM
deviations [mm] and corrector kicks [mrad] have to be discovered.
Any non-static, not pre-configured retrieval interface to the most
important performance analysis tool of the facility --archived data--
requires  clues to signal names, meaning, functionality, dimensions,
attributes. 

With the RTDB template approach the channel attributes are hidden within
these files and not available to the DB.  Consequently, no DB based data
relations are available for correlations (comparable
dimensions), projections (user, expert, device responsible) or dependencies
(triggered by). There is no browsable common data source that allows to
coordinate consistency of configurations requiring channel attributes (that
go beyond simple and clear assignments to well defined device classes).
The device model is tailored for modelling applications. For the archiver
retrieval it is useless continuously causing errors and
inconsistencies.

\section{Solution Attempt}
In a kind of clean-up effort the existing fragments used for RTDB
generation will be put into a new scheme. Sacrificing the feasibility to
generate complex RTDB templates with a graphical editor, now device class
tables, templates and I/O software modules ({\em records}) will be put
together as a new DB core. In view of the extendibility for high level
applications a purely DB based approach has been taken --- even though for
RTDB generation the framework of XML seems to be an attractive and well
suited alternative.

A new entity {\em gadget} is introduced connecting the data spaces {\em
Device}, {\em Channel} and {\em IOstruct}. These DB sections have to be set
up consequently in 3rd normal form, i.e. all device class properties,
channel attributes and I/O specifics have to be modelled in data and
relations, not in table structures. In blue-print the resulting data model
looks abstract and (intimidating) complex, but promising with respect to DB
flexibility, extendibility and cleanness. Expectation that today's
complicated configuration scripts collapse to simple sequences of DB
queries seem to be justified.


\section{SUMMARY}

The most difficult problem of a comprehensive configuration management
system is the determination of `natural' data structures and breaking them
down into localised data areas, dependencies and relations.  For a
relatively small installation like BESSY II the expected reward in
stability and configuration consistency does not clearly justify effort and
risk of the envisaged re-engineering task. One could argue that the present
mixture of DB based and hand edited configuration is an efficient
compromise. On the other hand the persistent and boring work-load due to
manual configuration update requirements is a constant source of
motivation to persue this reengineering project.

\end{document}